\documentclass[aps, twocolumn]{revtex4}

\usepackage{amsmath,bbm}
\usepackage{amssymb}
\usepackage{mathrsfs}
\usepackage{mathtools}
\usepackage{amsthm}
\usepackage{xcolor}
\usepackage{verbatim}
\usepackage{bm}
\usepackage{subfigure}
\usepackage[pdftex]{graphicx}
\usepackage{hyperref}
\hypersetup{
	colorlinks=true,
	linkcolor=blue,
	filecolor=magenta,      
	urlcolor=cyan,
}

\newcommand{\be}{\nopagebreak[3]\begin{equation}}
\newcommand{\ee}{\end{equation}}
\newcommand{\ba}{\nopagebreak[3]\begin{eqnarray}}
\newcommand{\ea}{\end{eqnarray}}
\DeclareFontFamily{U}{rsfs}{}         
\DeclareFontShape{U}{rsfs}{m}{n}{<5> rsfs5 <6><7> rsfs7          %
  <8><9><10><10.95><12><14.4><17.28><20.74><24.88> rsfs10}{}     %
\DeclareMathAlphabet{\mathfs}{U}{rsfs}{m}{n}                     %
\newcommand{\mfs}[1]{\mathfs {#1}}                               %
\newcommand{\n}{{\nonumber}}

\newcommand{\sD}{{\mfs D}}

\newcommand{\sT}{{\mfs T}}

\newcommand{\sM}{{\mfs M}}

\newcommand{\sO}{{\mfs O}}

 \usepackage{braket}

\usepackage{braket}

\begin{document}

\title{Renormalization of the  Quantum Stress Tensor Fluctuations and the Limits of Semiclassical Gravity}

\author{Alejandro Perez}
\email{perez@cpt.univ-mrs.fr}
\affiliation{{Aix-Marseille Universit\'e, Universit\'e de Toulon, CNRS, CPT, Marseille, France}}

\author{Daniel Sudarsky}
\email{sudarsky@nucleares.unam.mx}
\affiliation{Instituto de Ciencias Nucleares,
Universidad Nacional Aut\'onoma de M\'exico, M\'exico D.F. 04510, M\'exico}
\affiliation{Departament de Física Quantica i Astrofísica, Universidad de Barcelona, España}

\date{\today}


\begin{abstract} 
We analyze the expectation value of the energy--momentum tensor and its fluctuations in quantum field theory on curved spacetimes 
$\langle T_{ab} \rangle$. A generally accepeted condition for the conceptual consistency of semiclassical gravity, where $\langle T_{ab} \rangle$ represent the sources of the Einstein equations, is that the fluctuations of the energy momentum tensor remain small compared to its expectation value. We study the renormalization of both the energy--momentum tensor $\langle T_{ab}(x)\rangle_{\rm ren}$ and the fluctuation tensor $\langle T_{ab}(x) T_{cd}(x) \rangle_{\rm ren}$ for suitable Hadamard states, using the operator product expansion for a free scalar field on a fixed curved background. We show that states (usually referred to as `squeezed vacua')---arising naturally in black hole evaporation and in inflationary cosmology---fail to satisfy the natural semiclassicality  criterion.
\end{abstract}

\maketitle
In the absence of a consistent and well-defined theory of quantum gravity, semiclassical gravity—defined by Einstein’s equations coupled to the expectation value of the energy–momentum tensor of matter in a quantum state—is expected to provide useful insights into phenomena at the interface between gravity and quantum physics. Even though it is not entirely clear whether the dynamical equations of semiclassical gravity can be given a fully consistent interpretation---owing to questions about well-posedness in generic situations stemming from their higher-derivative character \cite{Flanagan:1996gw} {(see, \cite{Juarez-Aubry:2022qdp} for an approach  that appears as a promising path to deal with  those  issues,  and  including the initial value  formulation of the  theory)} ---the semiclassical framework nonetheless plays a central role in conceptual discussions. These range from the physics of black-hole evaporation via Hawking radiation, to analyses of the generalized second law, the fate of information, and various issues related to the emergence of structure in cosmology.

In order to  explore the regime of applicability of semiclassical gravity, it is natural to consider the quantum uncertainties of the energy–momentum tensor in comparison to the expectation value of that same quantity. The issue would be straightforward were it not for the fact that both the expectation value and the fluctuations are divergent notions that require suitable renormalization in order to yield finite and meaningful results.

The renormalization of expectation values of operators that are quadratic in the fundamental fields and their derivatives (such as the energy–momentum tensor of free fields) is customarily based on the Hadamard subtraction.   An axiomatic   analysis proposed by Wald \cite{Wald:1995yp} assures that  the recipe  is  unique up to the higher order curvature terms which are naturally expected to be suppressed by appropriate  powers of the Planck scale.

For operators built from higher powers of the fundamental fields and their derivatives (as in the case of the tensor product of the energy–momentum tensor with itself), subtracting a state-independent divergent quantity is no longer sufficient to eliminate ultraviolet divergences. In such cases, the more general framework of the operator product expansion (OPE)  { appears as the consistent path to renormalization}  \cite{WilsonZimmermann1972}.   In  \cite{Hollands:2023txn} its  generalization to arbitrary  (globally hyperbolic) curved spacetimes is proposed    and provides  a systematic and general procedure which might even  be  taken  as a more fundamental definition of the  quantum field theory itself.

{ The evaluation of the (renormalized) fluctuations of the energy–momentum tensor, and their relevance for assessing the validity of semiclassical gravity, has been considered in the past, beginning with the work of Ford \cite{FORD1982238} and later by Ford and Kuo \cite{Kuo:1993if}, where the problem was studied for quantum fields on Minkowski spacetime. Explicit calculations for certain curved spacetimes admitting a Euclidean continuation were subsequently carried out by Phillips and Hu \cite{Phillips:1996em} as well as in cosmological settings (see \cite{Hu:2008rga} and references therein). In these works, several perspectives and definitions of energy–momentum fluctuations were proposed, often invoking non-covariant expressions formulated in terms of specific mean values of components of the energy–momentum tensor, such as the “energy density.” In the early debates, much of the discussion centered on the choice of appropriate criteria for the validity of semiclassical gravity, motivated by the requirement that Minkowski spacetime endowed with the Poincaré-invariant vacuum should constitute a consistent solution of the corresponding field equations.

While our results agree in general lines with these earlier works and in the specific settings they considered, our approach offers a systematic and fully covariant framework that is, in principle, applicable to general spacetimes and general {(suitably tamed \cite{Hollands:2009bke})} states. Moreover, the OPE perspective ensures automatically that flat spacetime with matter fields in the Minkowski vacuum state is a consistent solution of semiclassical gravity (see \eqref{cricri}).

The results presented here should be  extendable to general quantum field theories. For simplicity, we will focus our discussion on the theory of a massive free scalar field described by the action
 \be
 S = \frac12 \int \sqrt{-g} d^4 x [\nabla^a \phi \nabla_a \phi  - m^2 \phi^2 ].
 \ee

As fields are operator-valued distributions in quantum field theory, the product of fields at the same point is ill-defined. The heuristic idea behind the renormalization of quadratic products begins by introducing a point-splitting regularization, in which one considers $\phi(x)\phi(x')$ for nearby spacetime events $x$ and $x'$, and then restricts attention to states that share a universal UV behavior (the so-called Hadamard states) in the coincidence limit. In this way one can separate the UV and IR pieces of the product as
\be\label{daniel}
{\phi(x)\phi(x')} \approx  C(x,x')_{\rm uv} \bf{1}  + {(\phi(x)\phi(x'))_{\rm ren}},
\ee
where  $\approx$ denotes equality in the limit $x \to x'$ when the expression is used to compute expectation values in Hadamard states. Here, $ C(x,x')_{\rm uv}$ captures the universal divergent behaviour in the coincidence limit, and $(\phi(x)\phi(x'))_{\rm ren}$ finite contributions. 

The previous heuristic discussion about the separation between 
UV divergent contributions and finite (physically relevant) long distance trends is at the conceptual heart of 
the rigorous Hadamard  subtraction procedure 
 usually employed to  compute  the  renormalized  expectation value of quadratic operators,  which in the case of the energy momentum tensor of a free field leads to the  expression  
 \be \label{renphiEMT} 
   \langle \psi | T_{ab} (x))  | \psi \rangle_{\rm ren}   =  \lim_{x  \to x' } {\sT}_{ab'}   [  \langle \psi |    \phi(x) \phi(x')  | \psi \rangle - H (x,  x')     ] \n 
  \ee
 where  the differential operator ${\sT}_{ab'}$  acting at the tangent planes of the  2-points is  given by
     \ba \label{renphiEMT} 
 {\sT}_{ab'}
  &=&   \nabla_a \nabla_{b'} - \frac{1}{2} g_{ab'}g^{cd'} \nabla_c \nabla_{d'}  - \frac{1}{2} g_{ab'} m^2  
\ea
with $g_{a{b'}}$ representing the parallel transport operators along the unique geodesic connecting the points $x$ and $x'$ within a common convex neighbourhood, and $H(x, x')$---which in equation \eqref{daniel} corresponds to $C(x,x')_{\rm uv}$---is the Hadamard bi-distributional kernel, constructed from geometrically defined quantities, which is state-independent and captures the UV behavior of the (Hadamard) states considered physically meaningful.  

The procedure admits the addition of a state independent divergence  free tensor  \cite{Wald:1995yp} whose relevant contributions correspond to a finite renormalization of the  cosmological constant and the Newton constant, in addition to two other contributions that could be obtained from variations of an action containing $R^2$ and $R_{ab} R^{ab}$ terms. Higher curvature contributions will be suppressed by a mass scale (naturally Planck's).

 When studying the fluctuations of the energy momentum tensor in a given quantum state of the scalar field, the strategy will be
to start with the 4-point function $\langle \psi |    \phi(x_1) \phi(x_2)  \phi(x_3) \phi(x_4)    | \psi \rangle$, from which we can define multi-tensorial quantity 
\be\label{gigo}
 \sT_{a_1b_2}\sT_{c_3,d_4} \langle \psi |    \phi(x_1) \phi(x_2)  \phi(x_3) \phi(x_4)    | \psi \rangle.
\ee
 The divergence structure of the 4-point function can be illustrated by considering the square of \eqref{daniel}, namely  
  \ba && (\phi(x) \phi(x'))^2 = C^{2}(x,x')_{\rm uv} {\bf 1}  +  \\ 
&&+ 2 C(x,x')_{\rm uv} (\phi(x)\phi(x'))_{\rm ren} +  ((\phi(x)\phi(x'))^2_{\rm ren}, \n 
   \ea
where one can identify a universal  UV  singular behavior in the first term, a second divergent term which is state dependent (due to the interference with the long distance trend \cite{Ford:2000vm, Ford2000StressTensor}), and the last one which is finite, and physically relevant for the renormalization of 
\eqref{gigo}.   The OPE is a systematic approach to extract such finite component which (in the coincidence limit) is defined as  $\braket{\psi|T_{ab}(x)T_{cd}(x)|\psi}_{\rm ren}$.

{ One might be concerned  about the ordering issues in taking the coincident limit and other ambiguities.  Fortunately, the OPE  axiomatic framework  ensures that the finite term is uniquely defined up to geometric (state independent) ambiguities similar to  those  occurring  in the renormalization of  $T_{ab} (x) $}.

{ The OPE  is very general and allows for the renormalization of the expectation value of arbitrary products of fields and their derivatives at a same point.  However, even  when the framework ensures  the success of the renormalization  program,  actual calculations can become  difficult in  concrete   practical  applications.

 Fortunately, in  case of free fields on curved backgrounds,  there is an alternative equivalent avenue for  computing  the renormalized  fluctuations of the energy momentum tensor.  With the OPE philosophy  in mind, we will find a shortcut using  the Wick  expansion of the  4-point function associated with a  specific  choice of vacuum state. This will facilitate explicit calculations on the one hand, and the interpretation of the result in situations where such vacua capture a concrete physical situation.} 

%
 
 In order to facilitate the interpretation of the results in situations where a preferred notion of vacuum state is available or physically motivated (as in the case of standard states such as the Hartle–Hawking state, the Unruh state, the in-state for static black hole backgrounds, or the adiabatic vacuum in cosmological settings), it is useful to relate the OPE formalism to the regularization procedure associated with normal ordering.
Even though the notion of normal ordering depends on the chosen vacuum state and is therefore highly ambiguous, the quantity
$\braket{\psi|T_{ab}(x)T_{cd}(x)|\psi}_{\rm ren}$
will be independent of the normal ordering chosen.

 Defining 
\be
 \braket{0|\phi(x_1)\phi(x_2)|0}_{\rm ren}\equiv \braket{0|\phi(x_1)\phi(x_2)|0}-H(x_1,x_2),
 \ee   
 and 
\ba \label{fifi}&&\!\!\!\!\!\!\!\!\!\braket{\psi |    \phi(x_1) \phi(x_2)  \phi(x_3) \phi(x_4)    | \psi}_{\rm ren} \equiv\\
&&\!\!\!\!\!\!\!\!\!  \equiv    \braket{\psi| \!:\! \phi(x_1) \phi(x_2)     \phi(x_3) \phi(x_4) \!:\!   | \psi }+\n  \\
  &&\!\!\!\!\!\!\!\!\!  \braket{\psi|    \!:\! \phi(x_1) \phi(x_2) \!:\!|\psi}   \braket{0|\phi(x_3) \phi(x_4) |0}_{\rm ren}\!+ {\rm other \ pairings} \n \\
  &&\!\!\!\!\!\!\!\!\!  \braket{0|\phi(x_1) \phi(x_2) |0}_{\rm ren}  \braket{0|\phi(x_3) \phi(x_4) |0}_{\rm ren}+ {\rm other \ pairings},\n
\ea
we show, in appendix \ref{apodo}, that 
\ba\label{casi-ope}
  &&  \braket{\psi |    \phi(x_1) \phi(x_2)  \phi(x_3) \phi(x_4)    | \psi} = \\ 
  &&   = \braket{\psi |    \phi(x_1) \phi(x_2)  \phi(x_3) \phi(x_4)    | \psi}_{\rm ren}
   +\n\\
  &&  +\braket{\psi| \phi(x_1) \phi(x_2) |\psi}_{\rm ren} H(x_3,x_4)+{\rm other \ pairings}
   \n \\
  &&  +H(x_1,x_2) H(x_3,x_4) +{\rm other \ pairings}.\n 
    \ea
Note that this corresponds to the OPE of the 4-point function  of a free field on a curved background. This follows from performing the multiple Taylor expansion of the renormalized quantities (those with the subindex ``ren") in \eqref{casi-ope}---for example writing $\braket{\psi| \phi(x) \phi(x+y) |\psi}_{\rm ren}=\braket{\psi| \phi(x)^2 |\psi}_{\rm ren}+y^\mu  \braket{ \psi|\phi(x)\partial_\mu  \phi(x) |\psi}_{\rm ren}+\cdots$.
As each of the divergent terms in  \eqref{casi-ope} are canonical, the finite term \eqref{fifi} does not depend on the choice of the vacuum state used in the normal ordering. Furthermore,  as  the various   terms  exhibit   the  appropriate  degree of divergence,    the   arrangement of the terms correspond to the  OPE  and  thus allow identification of the  desired  renormalized expectation  value of the composite operator  with no additional ambiguities beyond those already present in the renormalization of   energy momentum tensor (see Appendix \ref{ope}).

The renormalized square of the energy momentum tensor can now be given a meaning as
\ba
&& \braket{\psi|T_{ab}(x)T_{cd}(x)|\psi}_{\rm ren}\n \\
&& =\!\!\!\!\!\!\! \lim_{x_1,x_2,x_3,x_4\to x} \!\!\!\!\!\!\! \sT_{a_1b_2}\sT_{c_3d_4}\braket{\psi |    \phi(x_1) \phi(x_2)  \phi(x_3) \phi(x_4)    | \psi}_{\rm ren} \n
\ea
We can simplify the discussion of the fluctuations of the energy momentum tensor 
by focusing on an extreme idealization of a massive scalar field in its large mass limit. In that case
\be
\braket{T_{ab}}_{\rm ren }\approx  m^2\braket{\phi^2}_{\rm ren} g_{ab}. 
\ee   
Focusing on the trace, \eqref{fifi} implies that for any squeezed state (defined as a Bogoliubov transformation of an arbitrary vacuum state; see Appendix \ref{Squezzed} for more discussion) such that $\braket{\phi^2}_{\rm ren}\not=0$, we have
\be\label{traza}
{\Delta T^2_{\rm ren}}\equiv {\braket{T^2}_{\rm ren }-\braket{T}^2_{\rm ren }}\approx 2 {\braket{T}_{\rm ren}^2}.
\ee  
In what follows, we will show that the same result holds in full generality within a qualitatively realistic, yet simplified, 2-dimensional model.

\section{Explicit calculation in 2d}\label{2d}

The renormalized expectation value of the energy momentum tensor of a conformally invariant scalar field in 2d backgrounds can be
written as
\ba\label{previous}
&& \braket{\psi|T_{\pm \pm}(x)|\psi}_{\rm ren}=\lim_{x_1,x_2\to x}\braket{\psi|\partial_x^\pm\phi(x_1)\partial_x^\pm\phi(x_2) |\psi}_{\rm ren}\n  \\ &&\equiv \lim_{x_1,x_2\to x}\frac{\partial X^{\pm}_1}{\partial x^\pm_1}\frac{\partial X^\pm_2}{\partial x^\pm_2}\braket{\psi|: \partial_X^\pm\phi(x_1)\partial_X^\pm\phi(x_2):_X|\psi}\n,
\ea
where we denote by $:_X$ the normal ordered prescription based on the vacuum notion constructed out of the positive frequency notion associated with freely falling null coordinates $X^\pm$. The previous prescription is covariant by construction and geometric (see Section \ref{OPE}). Another advantage, that is only manifest in 2d, is that the Fock representations for arbitrary conformal null coordinates $x^\pm$---and their associated notion of positive frequency and associated vacuum $\ket{0_x}$---is readily accessible. In particular one can show that (see Appendix \ref{app})
\be\label{quehora}
\braket{\psi|T_{\pm \pm}(x)|\psi}_{\rm ren}=\braket{\psi|:T_{\pm \pm}:_x|\psi}-\frac{\hbar}{12\pi} \tau_{\pm}(\Omega),
\ee
where $:\, :_x$ denotes normal ordering constructed in terms of the vacuum $\ket{0_x}$, 
\be
\tau_{\pm}(\Omega)\equiv \left(\partial_{\pm}\log(\Omega)\partial_{\pm}\log(\Omega)-\partial^2_{\pm}\log(\Omega)\right),
\ee
with $\Omega$  the conformal factor appearing in the metric 
\be\label{metric}
ds^2=-\Omega^2 dx^+dx^-.
\ee
 One can show that the left hand side of \eqref{quehora} transforms covariantly under diffeomorphisms (with each term on the right transforming with the Schwarzian derivative with opposite signs). 
The conformal anomaly (see Appendix) fixes 
\be
\braket{\psi|T_{+ -}|\psi}=-\frac{\hbar}{12\pi} \partial_{+}\partial_{-}\log(\Omega).
\ee
For the fluctuations of $T_{ab}$ we proceed along similar lines, defining their renormalized value 
as the normal-ordered value with respect to the local freely falling vacuum, namely
\ba\label{titata}
&& \braket{\psi|T_{\pm \pm}(x)T_{\pm \pm}(x)|\psi}_{\rm ren}=  \\
&& \equiv \lim_{x_1,x_2,x_3,x_4\to x}\frac{\partial X^{\pm}_1}{\partial x^\pm_1}\frac{\partial X^\pm_2}{\partial x^\pm_2}\frac{\partial X^\pm_3}{\partial x^\pm_3}\frac{\partial X^\pm_4}{\partial x^\pm_4} \times \n \\ &&\times \braket{\psi|: \partial_X^\pm\phi(x_1)\partial_X^\pm\phi(x_2)\partial_X^\pm\phi(x_3)\partial_X^\pm\phi(x_4):_X|\psi}\n,
\ea
where, as in \eqref{previous}, we denote by $:_X$ the normal ordered prescription based on the vacuum notion constructed out of the positive frequency notion associated with freely falling  null coordinates $X^\pm$.
In  Appendix \ref{app} we explicitly show that the previous procedure  is equivalent to the
OPE perspective advocated in \cite{Hollands:2023txn}. We also show that the relevant renormalized fluctuations are given by (see \eqref{ckey})\ba\label{fifteen}
&& \braket{\psi|T_{\pm \pm}T_{\pm \pm}|\psi}_{\rm ren}\equiv
\braket{\psi|:T_{\pm \pm}T_{\pm \pm}:|\psi} 
\\ &&
+\frac{\hbar}{2\pi}\braket{\psi|:T_{\pm \pm}:|\psi}\tau_\pm(\Omega) + \frac{\hbar^2}{48 \pi^2}\tau_\pm^2(\Omega),\n 
\ea
whose covariance is granted by construction from the definition \eqref{titata}. The same method allows to calculate the $n$-point 
functions of $\partial_\pm \phi$ in a closed manner. 

We can now apply the previous formulae in order to compute the  renormalized fluctuations of the energy momentum tensor. We define
\be
\Delta T_{\pm\pm}^2(\ket{\psi})_{\rm ren}\equiv\! \braket{\psi|T_{\pm\pm}T_{\pm\pm}|\psi}_{\rm ren}\!-\braket{\psi|T_{\pm\pm}|\psi}_{\rm ren}^2.
\ee
For a coherent state $\ket{\phi_{\rm cl}}$, constructed by the displacement of the vacuum state used in the definition of the notion of normal ordering in our formulae, one has that
$\braket{\phi_{\rm cl}|:F(\phi):|\phi_{\rm cl}}=F(\phi_{\rm cl})$. Thus, using \eqref{quehora} and \eqref{fifteen} the condition for semiclassicality
${\Delta T_{\pm\pm}^2(\ket{\phi_{\rm cl}})_{\rm ren}}<{\braket{\phi_{\rm cl}|T_{\pm\pm}|\phi_{\rm cl}}^2_{\rm ren}}$ becomes
\ba\label{condition}
\frac{\hbar (5+\sqrt{26})}{12\pi}\tau_{\pm}<T_{\pm\pm}(\phi_{\rm cl})
\ea
when $\tau_{\pm}>0$ and 
\ba\label{condition}
\frac{\hbar (5-\sqrt{26})}{12\pi}\tau_{\pm}<T_{\pm\pm}(\phi_{\rm cl})
\ea
when  $\tau_{\pm}<0$ (recall that $T_{\pm\pm}(\phi_{\rm cl})\ge 0$).  Due to the simplicity of conformal coordinate transformations the previous are a covariant criteria.
Note that for any vacuum or squeezed vacuum state $\ket{0_x}$
\be\label{tota}
{{\Delta T_{\pm\pm}^2(\ket{0_x})_{\rm ren}}=2 {\braket{0_x|T_{\pm\pm}|0_x}^2_{\rm ren}}.}
\ee
which implies that such vacua cannot be considered semiclassical states. 

{The 2d simplified setting, while qualitatively realistic,  facilitates also the definition of a
coordinate independent semiclassicality criterion such as \eqref{condition}. 
In four dimensions,  however,  covariance would require a more careful treatment. One could for instance demand that  for all  sets of orthonormal  tetrads  $\lbrace e^a_I \rbrace  $ the quantities  $T_{IJ} \equiv e^a_I   e^b_J \hat T_{ab} $   would have  quantum uncertainties that  are small  compared  with the corresponding expectation values.  A   less restrictive  criteria  would  demand  only that  certain  scalar objects  such as  say  the  trace  $ T =  g^{ab}\hat T_{ab}$ (as in \eqref{traza})  be   such that its  expectation value  dominates over the corresponding uncertainties. 

In view of the previous discussion we propose as a precise criterion for the validity of semiclassical gravity at the spacetime point $x$ is that
\be\label{cricri}
\left|{\Delta T_{IJ}^2(x)}_{\rm ren}\right|\le \epsilon \braket{T_{IJ}(x)}_{\rm ren }^2
\ee
with $\epsilon<1$ for any tetrad  \footnote{One might naively worry about the fact that ambiguities in the renormalization of the quantities involved in the previous discussion might make the  criterion \eqref{cricri} ill defined.  Such ambiguous contributions appear  as  higher order modifications of  Einstein's equations  of the form $$G_{ab}= \ell^2_{p} \braket{\tilde{T}_{ab}(x)}_{\rm ren } +  \ell^2_{p}   (  \alpha \nabla_ a \nabla_b {R}   + \cdots) .$$
The natural suppression of these  terms  is ensured by the requirement that both curvatures and their rate of change be small compared with $\ell_p$.}.  Note that Minkowski with the Poincare invariant vacuum satisfies this at every point.

\section{Discussion}

In the context of 2d black holes,  states for which \eqref{tota} applies include emblematic cases such as the Boulware vacuum,  
the Hartle-Hawking state, the Unruh state, as well as the in-state (representing the particle creation in the setting of gravitational collapse).
This could have been naively anticipated from the Gaussian nature of such states as a general consequence of the Wick theorem. Here we have shown that the expectation remains correct after renormalization.  We note that in 4d the field $\Phi$ in  the spherically symmetric modes satisfies the Klein-Gordon equation (for a massive or massless cases) at the black hole horizon if $\Phi=\phi/r$ with $\phi$ satisfying the 
conformally invariant 2d equation of the field in the previous section. Moreover, the spherical mode contribution to the energy momentum tensor
at the black hole horizon, for the relevant component encoding the radiation flux across the horizon, is $T^{\rm sph}_{vv}=T^{\rm 2d}_{vv}/r^2$ (see Appendix \ref{apalache}).
Even when the in and Unruh quantum states would change in the generalization to 4d due to the backscattering by the effective potential that
is non trivial outside of the horizon, one expects that the previous identification would be qualitatively sensible for the expectation values and fluctuations of the above quantity.  This fact, combined with 
our discussion of the  4-point function OPE in 4d,  indicate that the  conclusions should remain valid in any dimension (even when explicit evaluation might be difficult).

We see that even a tiny deviation from the exact (Poincaré-invariant) Minkowski vacuum $\ket{0}_M$—where fluctuations are set to zero by the renormalization prescription—can produce large relative fluctuations of the energy–momentum tensor. This holds in particular for states obtained through a squeezing Bogoliubov transformation from $\ket{0}_M$, whether induced by curvature or simply by choice. Even when the condition \eqref{condition} may be satisfied locally in regions where matter fields are sufficiently excited, there will generically be `empty' regions in which relative fluctuations become significant. Of course, the gravitational effect of the fluctuating energy–momentum tensor will be of order $\hbar \tau_\pm$ in those regions, which in low-curvature regimes is expected to be very small compared to the local curvature scale \footnote{Near a black hole horizon of mass $M$, the ratio of curvature fluctuations to the background curvature is expected to be of order $m_p^2/M^2$.}.

Even when weak, their quantum gravitational effects can be significant in situations of interest, such as in the vicinity of macroscopic black holes. We show that the fluctuations of the energy–momentum tensor are of the same order of magnitude as the expectation values that drive black hole evaporation. This raises important questions about the validity of the semiclassical heuristic picture in analyzing the long-term evolution of such systems. In particular, our analysis seems to align with the insightful physical argument of Page \cite{Page:1979tc}, which implies large uncertainties in the black hole's position during evaporation, a conclusion strongly at odds with the semiclassical description.

The idea that semiclassical gravity cannot be applied in general situations is usually associated with scenarios involving macroscopic superpositions of gravitational sources at different locations (an idea that could, in principle, admit experimental confirmation \cite{Page:1981aj}). Here we show that, after proper renormalization, large fluctuations in gravitational sources are expected to be present in a broad class of quantum states of matter, including familiar and emblematic physical idealizations. This raises important questions regarding the status of many conceptual claims about the nature of quantum effects in the contexts of black-hole physics and primordial cosmology. It also reinforces the view that some of these issues can be properly addressed only within a framework in which the quantum nature of the gravitational degrees of freedom is fully taken into account.  
 
\section{Acknowledgments}

We are grateful to B. Kay,  B. Juarez Aubry for clarifying discussions in early stages of this work, and to I. Agullo, L. Freidel, M. Knecht, and E. Verdaguer for helpful exchanges. DS  acknowledges support from    the Proyect PAPIIT  No IG100124,  and   a sabbatical year Fellowship
 from UNAM,  through a  PASPA/DGAPA  and the hospitality of the Centre de Physique Th\'eorique in Marseille, and the  Departament de Física Quantica i Astrofísica  Universitat de Barcelona. This work was made possible through the support of the WOST, WithOut SpaceTime project (https://withoutspacetime.org), supported by Grant ID63683 from the John Templeton Foundation (JTF). The opinions expressed in this work are those of the author(s) and do not necessarily reflect the views of the John Templeton Foundation.

%
\providecommand{\href}[2]{#2}\begingroup\raggedright\endgroup

\clearpage
\begin{appendix}
\onecolumngrid

 \section{The OPE   and the uncertainties in the energy momentum tensor}\label{ope}

 In  order to  evaluate the uncertainties of the  energy  momentum  tensor we must compute the quantity $ \braket{\psi| T_{ab} (x)   T_{cd} (x) |\psi}_{\rm ren}   $.  We approach this  task  by considering the  OPE philosophy and focus  on the coincidence limit  of  the quantity  $ \braket{\psi| T_{ab} (x_1)   T_{cd} (x_2) |\psi}_{\rm ren} $. Here we  give a brief  description of the  formalism  underlying the  procedure which is  based on \cite{Hollands:2009bke}, and  applied to the  specific  case of the energy momentum tensor  for a free  scalar field   in  4 dimensions.  
 The OPE in this case yields\ba
{ T}_{a_1b_1} (x_1)   { T }_{c_2 d_2}(x_2)   = C^{(8)}_{a_1b_1,c_2d_2} ( x_1, x_2, y)  {1} +
\sum_{i=1}^{\infty}   C^{a^1_y a^2_y \cdots a_y^i; (8-i)}_{a_1b_1 , c_2 d_2} ( x_1,  x_2, y)  { O}^{(i)}_{a^1_y a^2_y \cdots a_y^i} (y), 
\ea
where $C^{a^1_y a^2_y \cdots a_y^i; (8-2i)}_{a_1b_1 , c_2 d_2} ( x_1,  x_2, y)$ are three-tensorial distributions in the tangent planes at $x_1$, $x_2$, and $y$ with 
abstract indices $a_1, b_1$, $a_2, b_2$ and $a^1_y a^2_y \cdots a_y^i$ in the three corresponding tangent planes, and $O^{(d)}_{\cdots}$ denotes operators with scaling dimension $d$.  For example the supra-index $(8-i)$ and $(i)$ denote the scaling dimension of the tensor valued distributions and operators in appearing in the OPE, which coincide with the power counting dimension in the present case as we are dealing with free fields (no anomalous dimensions involved).  Thus coefficients scale as $1/|\sigma|^{(8-i)}$ where $\sigma$ is the geodesic distance between the points (see \cite{Hollands:2009bke} for more details).  
 The OPE is  meant to be used in  evaluating expectation values calculated with the  physical states of the theory (see \cite{Hollands:2009bke}  for details) and for points as they approach coincidence.
One can write the first few terms with positive scaling dimension, namely
\ba\label{dada}
 && {T}_{a_1b_1} (x_1)   {T }_{a_2 b_2}(x_2)  =\\ && = C^{(8)}_{a_1b_1,c_2d_2} ( x_1, x_2, y)  {\bf   1} 
+ C^{a_y b_y; (4)}_{a_1b_1 , a_2 b_2} ( x_1,  x_2, y)  { O}^{(4)}_{a_y b_y} (y) + C^{a_y b_y c_y d_y; (0)}_{a_1b_1 , a_2 b_2} ( x_1,  x_2, y)  { O}^{(8)}_{a_y b_y c_y d_y} (y)+\cdots\n 
\ea
where the $\cdots$ denote all negative dimension contributions which vanish in the coincidence limit. 
The coefficients with scaling dimension $6$ and $2$ vanish due to the impossibility of having an intrinsically defined
four vector in terms of the available geometric data at this scaling dimension (e.g. one could take for instance $v_a=\nabla_aR$ but this is already dimension 3).
For suitable states $\ket{\psi}$ the expectation value of the third term in the previous equation is finite in the coincidence limit. Following the logic used in \cite{Hollands:2023txn}  for the renormalization of $\phi(x)^2$, we identify the renormalized value
\be\label{rerenono}\braket{\psi|{  T}_{a_1b_1} (x)   { T }_{a_2 b_2}(x)|\psi}_{\rm ren}\equiv \lim_{x_1,x_2\to x}\, C^{a_y b_y c_y d_y; (0)}_{a_1b_1 , a_2 b_2} ( x_1,  x_2, x) \braket{\psi| {O}^{(8)}_{a_y b_y c_y d_y} (x)|\psi} \ee
Thus, the OPE provides  a  recipe for the definition of  the quantity of interest,  namely the renormalized  expectation value,  in a suitable  physical state of the theory (i.e. the states for which the OPE  is  valid),  of the composite operator   $ T_{ab} (x)   T_{cd} (x) $,  just as the  Hadamard  substraction  provides a definition of the  renormalized  expectation value for a Hadamard   state   of the composite operator   $ T_{ab} (x)  $.

The axioms of the OPE imply that one can obtain \eqref{rerenono}
 by starting from the OPE corresponding to the product of four scalar field operators. 
 More precisely,  as algebraic  (and by considering   suitable limits  also  differential) relations  are required to be preserved in the corresponding expansion,  thus  we could  have  focused on the four point function, instead of looking at  the  energy momentum tensor directly,   and  would then have arrived  at  the desired  expression using its corresponding   OPE  evaluated on the state of interest (as  in  eq. (\ref{casi-ope})). Explicitly,  starting from \ba
&& \phi(x_1)\phi(x_2)\phi(x_3)\phi(x_4)=C^{(4)}(x_1,x_2,x_3,x_4; y) {\bf  \hat 1} + C^{(2)} (x_1,x_2,x_3,x_4; y) { \phi^2}(y)+ C^{(0)}(x_1,x_2,x_3,x_4; y)  \phi^4(y)+\n \\ 
&& + C^{ab; (0)}(x_1,x_2,x_3,x_4; y) \partial_a\phi(y)\partial_b\phi(y) + C^{ab; (-2)}(x_1,x_2,x_3,x_4; y)\phi^2(y) \partial_a\phi(y)\partial_b\phi(y)\n \\
&& + C^{abcd; (-4)}(x_1,x_2,x_3,x_4; y) \partial_a\phi(y)\partial_b\phi(y)\partial_c\phi(y)\partial_d\phi(y) +\cdots ,
\ea
one has that 
\ba
&& \braket{\psi|{T}_{a_1b_2} (x)   {  T }_{a_3 b_4}(x)|\psi}_{\rm ren}
=\lim_{x_1,x_2, x_3, x_4\to x}\,  \underbrace{\left(\sT_{a_1b_2}  \sT_{a_3b_4} \braket{\psi|\phi(x)\phi(x)\phi(x)\phi(x)|\psi}\right)}_{d=8\ \rm term \ in \ the\  OPE\  expansion} ,\n 
\ea
with $\sT_{ab'}$ defined in \eqref{renphiEMT}. Note that $C^{(0)}(x_1,x_2,x_3,x_4; y)$, $C^{ab; (-2)}(x_1,x_2,x_3,x_4; y)$, and $C^{abcd; (-4)}(x_1,x_2,x_3,x_4; y)$ contribute to the final result.  The  general expression in four dimensions can be expected to  be  rather cumbersome yet available given the four and two point correlation functions in the given states (as shown in equations \eqref{fifi} and \eqref{casi-ope}).  We illustrated in this work the general  approach explicitly using the simple case of a  massless  scalar file in  2  dimensions yet the scheme and the conclusions are general. Indeed, higher powers of the energy momentum tensor can be renormalized with the same framework.

Up to some subtle issues that we will not discuss here,
the requirement that  the l.h.s. of \eqref{dada} be  divergence-less on both points implies that each one of the  coefficients  on its own  must  be  divergence free in both points as well, as there is no general   possibility of cancellations  between the terms due to the independence of the operators  $ \hat O^{(i)} (y) $ occurring   in the expansion.  That property, in turn, suggests that the ambiguities occurring in the identification of the coefficients are closely related to, and appear not to exceed, those arising in the standard renormalization of  $\langle T_{ab} \rangle$ itself. 

A formal link between the renormalization ambiguities of the expectation value of the quantities of interest here and  the ambiguities 
in the effective action can be established via the path integral formulation.  The expectation value of the energy momentum tensor is given by
\be \braket{T_{ab}(x)}\equiv \frac{i\hbar}{Z \sqrt{|g(x)|}}\frac{\delta}{\delta g^{ab}(x)} \underbrace{\left( \int \sD g_{\mu\nu} \sD\phi \exp\left(\frac{i}{\hbar} S_M[g_{\mu\nu}, \phi]\right)\right)}_{\equiv Z}.
\ee
where $S_M[g_{\mu\nu}, \phi]$ is the matter action
defined as the full action minus the (bare) Einstein-Hilbert action, i. e. , it contains all the terms leading to possible higher curvature corrections to Einstein's equations (information about the quantum state is meant to be encoded in the boundary conditions of the functional integral). The renormalization ambiguities in the $\braket{T_{ab}(x)}$ mentioned in the introduction (that follow from Wald's axioms \cite{Wald:1995yp}) are thus connected to the renormalization of the parameters in $S_M[g_{\mu\nu}, \phi]$ like the Newton and cosmological constants as well as the higher curvature couplings like $\alpha \sqrt{|g|}R^2+\beta \sqrt{|g|} R_{ab}R^{ab}+\cdots$ appearing in the Lagrangian defining $S_M[g_{\mu\nu}, \phi]$. Similarly, we can study second variations from which we get
\ba && \frac{-\hbar^2}{Z \sqrt{|g(y)|}}\frac{\delta}{\delta g^{cd}(y)}\frac{\delta}{\sqrt{|g(x)|}\delta g^{ab}(x)} \left( \int \sD g_{\mu\nu} \sD\phi \exp\left(\frac{i}{\hbar} S_M[g_{\mu\nu}, \phi]\right)\right)= \\ \n
&&=\underbrace{\frac{1}{Z}\left( \int \sD g_{\mu\nu} \sD\phi\, T_{ab}(x)T_{cd}(y)\exp\left(\frac{i}{\hbar} S_M[g_{\mu\nu}, \phi]\right)\right)}_{\equiv  \braket{T_{ab}(x)T_{cd}(y)}}+\overbrace{\frac{i\hbar}{Z }\left( \int \sD g_{\mu\nu} \sD\phi\, \frac{\delta T_{ab}(x)}{\sqrt{|g(y)|}\delta g^{cd}(y)} \exp\left(\frac{i}{\hbar} S_M[g_{\mu\nu}, \phi]\right)\right)}^{{\rm contact\ term}=\delta(x,y) \braket{O_{abcd}(y)}+\cdots}.
\ea
This implies that---apart from singular contact terms that would not contribute to $\braket{T_{ab}(x)T_{cd}(x)}_{\rm ren}$---the finite operator expectation value $\braket{T_{ab}(x)T_{cd}(x)}_{\rm ren}$ shares the ambiguities with $\braket{T_{ab}(x)}_{\rm ren} \braket{T_{cd}(x)}_{\rm ren}$, which thus should cancel in the expression of $\Delta (T_{ab}T_{cd})(x)$. 

\section{Four point function in 4d}\label{apodo}

Using the Wick theorem and a choice of vacuum state $\ket{0}$,  the  4-point function \eqref{casi-ope} can be expressed as follows
 \ba
  && \braket{\psi |    \phi(x_1) \phi(x_2)  \phi(x_3) \phi(x_4)    | \psi} =
   \braket{\psi| \!:\! \phi(x_1) \phi(x_2)     \phi(x_3) \phi(x_4) \!:\!   | \psi }+\n \\
  &&  \braket{\psi|    \!:\! \phi(x_1) \phi(x_2) \!:\!|\psi}   \braket{0|\phi(x_3) \phi(x_4) |0}+ \braket{\psi|    \!:\! \phi(x_1) \phi(x_3) \!:\!|\psi}   \braket{0|\phi(x_2) \phi(x_4) |0}+ \braket{\psi|    \!:\! \phi(x_1) \phi(x_4) \!:\!|\psi}   \braket{0|\phi(x_2) \phi(x_3) |0} + 
  \n \\
  &&  \braket{\psi|    \!:\! \phi(x_2) \phi(x_3) \!:\!|\psi}   \braket{0|\phi(x_1) \phi(x_4) |0}+ \braket{\psi|    \!:\! \phi(x_2) \phi(x_4) \!:\!|\psi}   \braket{0|\phi(x_1) \phi(x_3) |0}+ \braket{\psi|    \!:\! \phi(x_3) \phi(x_4) \!:\!|\psi}   \braket{0|\phi(x_1) \phi(x_2) |0} + \n
   \n \\
  &&  \braket{0|  \phi(x_2) \phi(x_3)|0}   \braket{0|\phi(x_1) \phi(x_4) |0}+ \braket{0| \phi(x_2) \phi(x_4) |0}   \braket{0|\phi(x_1) \phi(x_3) |0}+ \braket{0|\phi(x_3) \phi(x_4)|0}   \braket{0|\phi(x_1) \phi(x_2) |0}. \n
    \ea
We assume that the reference vacuum state $\ket{0}$ is a Hadamard state and define the (finite in the coincidence limit) quantity
 \be
 \braket{0|\phi(x_1)\phi(x_2)|0}_{\rm ren}\equiv \braket{0|\phi(x_1)\phi(x_2)|0}-H(x_1,x_2).
 \ee   
 Using the previous equation in order to replace the vacuum expectation values in terms of renormalized  2-point function and the Hadamard bi distribution we get
\ba\label{A2}
  && \braket{\psi |    \phi(x_1) \phi(x_2)  \phi(x_3) \phi(x_4)    | \psi} =
   \braket{\psi| \!:\! \phi(x_1) \phi(x_2)     \phi(x_3) \phi(x_4) \!:\!   | \psi }+ \\
  &&+  \braket{\psi|    \!:\! \phi(x_1) \phi(x_2) \!:\!|\psi}   \braket{0|\phi(x_3) \phi(x_4) |0}_{\rm ren}+ \braket{\psi|    \!:\! \phi(x_1) \phi(x_3) \!:\!|\psi}   \braket{0|\phi(x_2) \phi(x_4) |0}_{\rm ren} +\n \\
&&  + \braket{\psi|    \!:\! \phi(x_1) \phi(x_4) \!:\!|\psi}   \braket{0|\phi(x_2) \phi(x_3) |0}_{\rm ren} +  \braket{\psi|    \!:\! \phi(x_2) \phi(x_3) \!:\!|\psi}   \braket{0|\phi(x_1) \phi(x_4) |0}_{\rm ren}+\n \\ && + \braket{\psi|    \!:\! \phi(x_2) \phi(x_4) \!:\!|\psi}   \braket{0|\phi(x_1) \phi(x_3) |0}_{\rm ren}+ \braket{\psi|    \!:\! \phi(x_3) \phi(x_4) \!:\!|\psi}   \braket{0|\phi(x_1) \phi(x_2) |0}_{\rm ren} + \n
   \n \\
  && + \braket{0|\phi(x_2) \phi(x_3) |0}_{\rm ren}\!   \braket{0|\phi(x_1) \phi(x_4) |0}_{\rm ren}+ \braket{0| \phi(x_2) \phi(x_4)|0}_{\rm ren} \!  \braket{0|\phi(x_1) \phi(x_3) |0}_{\rm ren}
 +\n \\ && 
  + \braket{0|   \phi(x_3) \phi(x_4) |0}_{\rm ren} \! \braket{0|\phi(x_1) \phi(x_2) |0}_{\rm ren} +\n\\
  && +  \braket{\psi|    \!:\! \phi(x_1) \phi(x_2) \!:\!|\psi}  H(x_3,x_4)+ \braket{\psi|    \!:\! \phi(x_1) \phi(x_3) \!:\!|\psi} H(x_2,x_4)  + \braket{\psi|    \!:\! \phi(x_1) \phi(x_4) \!:\!|\psi}  H(x_2,x_3) + 
  \n \\
  && + \braket{\psi|    \!:\! \phi(x_2) \phi(x_3) \!:\!|\psi}   H(x_1,x_4)+ \braket{\psi|    \!:\! \phi(x_2) \phi(x_4) \!:\!|\psi}   H(x_1,x_3)+ \braket{\psi|    \!:\! \phi(x_3) \phi(x_4) \!:\!|\psi} H(x_1,x_2) + \n
   \n \\
  &&+ H(x_2,x_3)   \braket{0|\phi(x_1) \phi(x_4) |0}_{\rm ren}+ H(x_2,x_4)  \braket{0|\phi(x_1) \phi(x_3) |0}_{\rm ren}+ H(x_3,x_4)   \braket{0|\phi(x_1) \phi(x_2) |0}_{\rm ren} \n \\
  &&+ H(x_1,x_2)   \braket{0|\phi(x_3) \phi(x_4) |0}_{\rm ren}+ H(x_1,x_3)  \braket{0|\phi(x_2) \phi(x_4) |0}_{\rm ren}+ H(x_1,x_4)   \braket{0|\phi(x_2) \phi(x_3) |0}_{\rm ren}+ \n \\
  &&+ H(x_1,x_2) H(x_3,x_4) +H(x_1,x_3) H(x_2,x_4) +H(x_1,x_4) H(x_2,x_3) .\n 
    \ea
Inspection of the previous equation allows for identifying the term that is free of UV divergences, namely
\ba \label{fifi-a}&&\braket{\psi |    \phi(x_1) \phi(x_2)  \phi(x_3) \phi(x_4)    | \psi}_{\rm ren} \equiv
   \braket{\psi| \!:\! \phi(x_1) \phi(x_2)     \phi(x_3) \phi(x_4) \!:\!   | \psi }+ \\
  &&+  \braket{\psi|    \!:\! \phi(x_1) \phi(x_2) \!:\!|\psi}   \braket{0|\phi(x_3) \phi(x_4) |0}_{\rm ren}+ \braket{\psi|    \!:\! \phi(x_1) \phi(x_3) \!:\!|\psi}   \braket{0|\phi(x_2) \phi(x_4) |0}_{\rm ren}+ \n \\ 
  &&+ \braket{\psi|    \!:\! \phi(x_1) \phi(x_4) \!:\!|\psi}   \braket{0|\phi(x_2) \phi(x_3) |0}_{\rm ren} + 
  \braket{\psi|    \!:\! \phi(x_2) \phi(x_3) \!:\!|\psi}   \braket{0|\phi(x_1) \phi(x_4) |0}_{\rm ren}+ \n \\
 &&+ \braket{\psi|    \!:\! \phi(x_2) \phi(x_4) \!:\!|\psi}   \braket{0|\phi(x_1) \phi(x_3) |0}_{\rm ren}+ \braket{\psi|    \!:\! \phi(x_3) \phi(x_4) \!:\!|\psi}   \braket{0|\phi(x_1) \phi(x_2) |0}_{\rm ren} + \n
   \n \\
  && + \braket{0|\phi(x_1) \phi(x_2) |0}_{\rm ren}  \braket{0|\phi(x_3) \phi(x_4) |0}_{\rm ren}+ \braket{0| \phi(x_1) \phi(x_3)|0}_{\rm ren}   \braket{0|\phi(x_2) \phi(x_4) |0}_{\rm ren} +\n \\ 
  &&+ \braket{0|   \phi(x_1) \phi(x_4) |0}_{\rm ren}   \braket{0|\phi(x_2) \phi(x_3) |0}_{\rm ren},\n
\ea
which, when replaced in \eqref{A2}, yields
    \ba\label{casi-ope-a}
  && \braket{\psi |    \phi(x_1) \phi(x_2)  \phi(x_3) \phi(x_4)    | \psi} = \braket{\psi |    \phi(x_1) \phi(x_2)  \phi(x_3) \phi(x_4)    | \psi}_{\rm ren}
   +\n\\
  &&+  (\braket{\psi|    \!:\! \phi(x_1) \phi(x_2) \!:\!|\psi} + \braket{0|\phi(x_1) \phi(x_2) |0}_{\rm ren}) H(x_3,x_4)+ \left(\braket{\psi|    \!:\! \phi(x_1) \phi(x_3) \!:\!|\psi} + \braket{0|\phi(x_1) \phi(x_3) |0}_{\rm ren}\right) H(x_2,x_4)  + \n \\ 
  &&+ (\braket{\psi|    \!:\! \phi(x_1) \phi(x_4) \!:\!|\psi} + \braket{0|\phi(x_1) \phi(x_4) |0}_{\rm ren}) H(x_2,x_3) + 
   (\braket{\psi|    \!:\! \phi(x_2) \phi(x_3) \!:\!|\psi} + \braket{0|\phi(x_2) \phi(x_3) |0}_{\rm ren})  H(x_1,x_4)+ \n \\ &&+
   (\braket{\psi|    \!:\! \phi(x_2) \phi(x_4) \!:\!|\psi}+ \braket{0|\phi(x_2) \phi(x_4) |0}_{\rm ren})   H(x_1,x_3)
   + (\braket{\psi|    \!:\! \phi(x_3) \phi(x_4) \!:\!|\psi}+ \braket{0|\phi(x_3) \phi(x_4) |0}_{\rm ren}) H(x_1,x_2) + \n
   \n \\
  &&+ H(x_1,x_2) H(x_3,x_4) +H(x_1,x_3) H(x_2,x_4) +H(x_1,x_4) H(x_2,x_3) .
    \ea
Despite the explicit appearance of normal-ordered contributions in the renormalized correlation function \eqref{fifi-a}, one can see that it is indeed independent of the choice of the reference vacuum state used to define normal ordering. This becomes clear by performing one final rearrangement of \eqref{casi-ope} as follows:
 \ba\label{casi-ope-a}
  && \braket{\psi |    \phi(x_1) \phi(x_2)  \phi(x_3) \phi(x_4)    | \psi} = \braket{\psi |    \phi(x_1) \phi(x_2)  \phi(x_3) \phi(x_4)    | \psi}_{\rm ren}
   +\n\\
  && +  (\braket{\psi| \phi(x_1) \phi(x_2) |\psi} - H(x_1,x_2)) H(x_2,x_3)+ \left(\braket{\psi| \phi(x_1) \phi(x_3) |\psi} -H(x_1,x_3) \right) H(x_2,x_4)  + \n \\ 
  && +(\braket{\psi|   \phi(x_1) \phi(x_4) |\psi} - H(x_1,x_4)) H(x_2,x_3) + 
   (\braket{\psi| \phi(x_2) \phi(x_3) |\psi} - H(x_2,x_3))  H(x_1,x_4)+ \n \\ && +
   (\braket{\psi| \phi(x_2) \phi(x_4) |\psi}- H(x_2,x_4))   H(x_1,x_3)
   + (\braket{\psi|\phi(x_3) \phi(x_4) |\psi}- H(x_3,x_4)) H(x_1,x_2) + \n
   \n \\
  && +H(x_1,x_2) H(x_3,x_4) +H(x_1,x_3) H(x_2,x_4) + H(x_1,x_4) H(x_2,x_3) .
    \ea
or
\ba\label{casi-ope-a}
  && \braket{\psi |    \phi(x_1) \phi(x_2)  \phi(x_3) \phi(x_4)    | \psi} = \braket{\psi |    \phi(x_1) \phi(x_2)  \phi(x_3) \phi(x_4)    | \psi}_{\rm ren}
   +\n\\
  &&  \braket{\psi| \phi(x_1) \phi(x_2) |\psi}_{\rm ren} H(x_2,x_3)+ \braket{\psi| \phi(x_1) \phi(x_3) |\psi}_{\rm ren} H(x_2,x_4)  + \n \\ 
  && \braket{\psi|   \phi(x_1) \phi(x_4) |\psi}_{\rm ren} H(x_2,x_3) + 
   \braket{\psi| \phi(x_2) \phi(x_3) |\psi}_{\rm ren}  H(x_1,x_4)+ \n \\ &&
   \braket{\psi| \phi(x_2) \phi(x_4) |\psi}_{\rm ren}   H(x_1,x_3)
   + \braket{\psi|\phi(x_3) \phi(x_4) |\psi}_{\rm ren} H(x_1,x_2) + \n
   \n \\
  && +H(x_1,x_2) H(x_3,x_4) +H(x_1,x_3) H(x_2,x_4) + H(x_1,x_4) H(x_2,x_3).
    \ea   
The left hand side is obviously independent of the choice of reference vacuum. All the terms 
on the right hand side, except from the first one, are also manifestly independent of the reference vacuum. Thus, the first term 
 $ \braket{\psi |    \phi(x_1) \phi(x_2)  \phi(x_3) \phi(x_4)    | \psi}_{\rm ren}$ is independent of the reference vacuum state used in the expression \eqref{fifi-a}.

\section{Renormalization details in a conformal 2d quantum field theory}\label{app}

The analysis of the previous section could be adapted directly to the computations performed here. However, because the two-point function of a massless scalar field $\phi$ is not well defined (due to infrared divergences) and due to special features associated with the conformal symmetry of the model, it is more convenient to carry out the renormalization of the energy–momentum tensor (and its fluctuations) directly in terms of the well-behaved and unambiguous correlation functions of $d\phi$ instead. Added interest of this section is that it starts with a self-contained,  concise, and explicit review of the covariant renormalization the expectation value $T_{\mu\nu}$ in 2d (more details of which can be found for instance in \cite{Fabbri:2005mw} and references therein). 

In 2d there is a Fock representation for each conformal coordinate system with the metric of the form (\ref{metric}) with the field operator written 
in the canonical form 
\be
\phi(x)=\int_{0}^\infty \frac{d\omega}{4\pi \omega} \left( a^+_\omega e^{-i\omega x^+}+ a^-_\omega e^{-i\omega x^-}+ h.c.\right), 
\ee
with $a^\pm_\omega$ the annihilation operators for right and left moving modes with frequency $\omega$.
The two-point function in the associated vacuum is IR divergent and thus ambiguous, yet it can be regularized by introducing an infrared cut-off scale $L$  and write 
\be
\braket{0_x|\phi(x_1)\phi(x_2)|0_x}\approx \frac{\hbar}{4\pi}\left(2\gamma+\log\left[\frac{|(x^+_1-x^+_2)(x^-_1-x^-_2)|}{L^2}\right]\right),
\ee
where $\approx$ indicates that the expression is an approximation whose corrections play no role in the formulas that follow—those involving derivatives of the above correlation function—in the coincidence limit. The cut-off dependence disappears when considering the  2-point function of $d\phi$ instead, which is the one of interest 
in the definition of the energy momentum tensor for the conformal scalar field. Namely,
\be
\braket{0_x|\partial_\pm\phi(x_1)\partial_\pm\phi(x_2)|0_x}\approx-\frac{\hbar}{12\pi} \frac{1}{|x^\pm_1-x^\pm_2|^2}.
\ee 
For further use we will denote respectively $\Delta^{\pm}\equiv x^\pm_1-x^\pm_2$ and will often simply use just $\Delta$ when the context is clear. 
On can regularize the expectation value of the energy momentum tensor by a point splitting procedure and the corresponding subtraction 
\ba
\braket{\psi|:T_{\pm \pm}(x):_x|\psi}&=&
\lim_{\Delta \to 0}\left[\braket{\psi|\partial_{\pm}\phi(x)\partial_{\pm}\phi(x+\Delta)|\psi}-\braket{0|\partial_{\pm}\phi(x)\partial_{\pm}\phi(x+\Delta)|0}\right]=\n \\
&=&\lim_{\Delta \to 0}\left[\braket{\psi|\partial_{\pm}\phi(x)\partial_{\pm}\phi(x+\Delta)|\psi}+\frac{\hbar}{4\pi} \frac{1}{\Delta^2}\right],
\ea
which in this case corresponds to simply normal ordering the creation and annihilation operators in the mode expansion of the basic fields.
This procedure yields a regular result in the coincidence limit for suitable states $\ket{\psi}$ with vacuum-like  behaviour in the UV (Hadamard states). The first issue associated with such a prescription is that it depends on the notion of vacuum chosen which is ambiguous. A geometric prescription is based on the Hadamard subtraction (see section \ref{OPE}) and the assumption that $\ket{\psi}$ is a Hadamard state (see for instance \cite{Wald:1995yp}).
Nevertheless, the previous (apparently non-geometric) prescription is of great practical use in 2d and leads (as we show here) to equivalent results. 

We notice that two normal ordering prescriptions are related by
\ba
\braket{\psi|:T_{\pm\pm}(x):_x|\psi}&-&\left(\frac{\partial y^\pm}{\partial x^\pm}\right)^2 \braket{\psi|:T_{\pm\pm}(y):_y|\psi}= \\
&=&\lim_{\Delta\to 0}\frac{\hbar}{4\pi}\left( \frac{1}{|\Delta^{\pm}|^2}-\frac{\frac{\partial y^\pm}{\partial x^\pm}(x) \frac{\partial y^\pm}{\partial x^\pm}(x+\Delta)}{|y^\pm(x)-y^\pm(x+\Delta)|^2}\right)=-\frac{\hbar}{24\pi} \{y^\pm,x^\pm\},\n
\ea
where 
\be \{y,x\}\equiv \frac{\frac{d^3 y}{dx^3}}{\frac{dy}{dx}}-\frac32 \left( \frac{\frac{d^2 y}{dx^2}}{\frac{dy}{dx}}\right)^2\ee
is the so-called Schwarzian derivative. 

\subsection{Calculation of $\braket{\psi|T_{\pm \pm}(x)|\psi}_{\rm ren}$}
One uses the previous relation between different notions of normal ordered regularizations in order to construct the covariant renormalization prescription \eqref{previous} which leads to
\be\label{kiko}
\braket{\psi|T_{\pm \pm}(x)|\psi}_{\rm ren}=\braket{\psi|:T_{\pm \pm}:_x|\psi}-\frac{\hbar}{12\pi} \left(\partial_{\pm}\log(\Omega)\partial_{\pm}\log(\Omega)-\partial^2_{\pm}\log(\Omega)\right),
\ee
which is equation \eqref{quehora} relating the renormalized expectation value to the normal ordered version (in an arbitrary vacuum $\ket{0_x}$) and the Schwarzian derivative
\be\label{swm}
\frac{\hbar}{24\pi}\{X^\pm,x^\pm\}=-\frac{\hbar}{12\pi}\left(\partial_{\pm}\log(\Omega)\partial_{\pm}\log(\Omega)-\partial^2_{\pm}\log(\Omega)\right)
\ee
 where $X^\pm$ are Riemann normal coordinates \cite{Fabbri:2005mw}
 \be
X^{\pm}_p\equiv \Omega (x^{\pm} -x^{\pm}(p))+\Omega \, \partial_{\pm} \log(\Omega) (x^{\pm} -x^{\pm}(p))^2+\Omega \left(\frac{1}{3}\partial_\pm^2(\log(\Omega))+\frac{2}{3} (\partial_\pm \log(\Omega))^2\right)(x^{\pm} -x^{\pm}(p))^3+\cdots
\ee
 with $\Omega$ is the conformal factor in \eqref{metric}. These coordinates are such that the metric \eqref{metric}
 takes the form \be ds^2=-dX^+dX^-\left(1+ \frac{1}{4}\Omega^4R (X^+-X^+(p))(X^- -X^-(p))\right)+\sO((X-X(p))^3), \ee
 in a local neighbourhood of every point $p \in \sM$, and where $R$ is the scalar curvature (see \eqref{b12} below).
 The classical expression of $T_{+-}$ vanishes identically due to conformal invariance.
 The divergence $\nabla^aT_{ab}=g^{ca}(\partial_cT_{ab}-\Gamma^d_{ca} T_{db})$ in the conformal coordinates  become $g^{+ -}(\partial_+T_{-b}+\partial_-T_{+b})$ as the only non vanishing Christoffel symbols $\Gamma_{\pm\pm}^\pm$ drop out due to $T_{+-}=0$. Using \eqref{kiko} we get
 \be
 \nabla^a\braket{\psi|T_{a\pm}|\psi}_{\rm ren}=2\Omega^{-2}\partial_{\mp} \braket{\psi|T_{\pm\pm}|\psi}=-\frac{\hbar}{12\pi} \partial_{\mp}\left(\partial_{\pm}\log(\Omega)\partial_{\pm}\log(\Omega)-\partial^2_{\pm}\log(\Omega)\right)=-\frac{\hbar}{48\pi} \partial_{\pm} R,
\ee
where we used that \cite{Wald:1984rg}
\be\label{b12}
R=-2g^{ab}\nabla_a\nabla_b\log{\Omega}=8\Omega^{-2} \partial_+\partial_-\log(\Omega).
\ee
The previous equation can be written covariantly as
\be
\nabla^a\braket{\psi|T_{ab}|\psi}_{\rm ren}=-\frac{\hbar}{48\pi} \nabla_b R,
\ee
which we call the diffeomorphism anomaly. One can absorb the anomaly defining a modified renormalization prescription via
\be
 \braket{\psi|\tilde T_{ab}|\psi}_{\rm ren}\equiv \braket{\psi|T_{ab}|\psi}_{\rm ren}+\frac{\hbar}{48\pi} R g_{ab}
\ee
which is divergence free, yet not traceless
\be
g^{ab}\braket{\psi|\tilde T_{ab}|\psi}_{\rm ren}=\frac{\hbar}{24\pi} R, 
\ee
which is the so-called trace anomaly. Both posibilities are also realized in 4d.  In the first case diffeomorphisms are broken down to volume-preserving diffeomorphism and the gravity field equations are those of semiclassical trace-free gravity (unimodular gravity). In the second case the gravity field equations are semiclassical Einsteins equations. Both alternatives give the same gravitational dynamics \cite{Josset:2016vrq, Amadei:2021aqd}. 

\subsection{Calculation of $\braket{\psi|T_{\pm \pm}T_{\pm \pm}|\psi}_{\rm ren}$}
For the  4-point function needed in the point-splitting-regularization of the square of the energy momentum tensor we shall introduce a simplifying notation where 
\ba
&& \braket{\psi|\partial_\pm\phi(x_1)\partial_\pm\phi(x_2)\partial_\pm\phi(x_3)\partial_\pm\phi(x_4)|\psi}\equiv\braket{1_\pm2_\pm3_\pm4_\pm}^x_\psi\n \\
&&\braket{\psi|:\partial_\pm\phi(x_1)\partial_\pm\phi(x_2)\partial_\pm\phi(x_3)\partial_\pm\phi(x_4):|\psi}\equiv \, \braket{:1_\pm2_\pm3_\pm4_\pm\!:}_\psi^x,
\ea
and the last line defines the normal ordering in the vacuum $\ket{0_x}$ corresponding to the conformal coordinates $x^\pm$. We also dropped all reference to the two cases $\pm$ as the algebra that follows is valid for both. The cross term  4-point correlation functions are all trivially related to the  2-point functions as the right and left moving field modes commute.
We will also denote the  2-point function 
as follows
\be
\braket{0_x|\partial_\pm\phi(x_1)\partial_\pm\phi(x_2)|0_x}=-\frac{\hbar}{12\pi} \frac{1}{|x^\pm_1-x^\pm_2|^2}\equiv \xi_{12}^{x^\pm}=\xi_{21}^{x^\pm}
\ee 
With this notation, the Wick theorem  implies that
\ba\label{fff}
&& \braket{1_\pm2_\pm3_\pm4_\pm}^x_\psi = \braket{:1_\pm2_\pm3_\pm4_\pm\!:}^x_\psi+\n \\
&& +\braket{:1_\pm 2_\pm\!:}^x_\psi \xi_{34}^{x^\pm}+\braket{:1_\pm 3_\pm\!:}^x_\psi \xi_{24}^{x^\pm}+\braket{:1_\pm 4_\pm\!:}^x_\psi\xi_{23}^{x^\pm}+\braket{:2_\pm 3_\pm\!:}^x_\psi \xi_{14}^{x^\pm}+\braket{:2_\pm 4_\pm\!:}^x_\psi \xi_{13}^{x^\pm}+\braket{:3_\pm 4_\pm\!:}^x_\psi \xi_{12}^{x^\pm}+\n \\
&&  +\xi_{12}^{x^\pm}  \xi_{34}^{x^\pm}+ \xi_{13}^{x^\pm}  \xi_{24}^{x^\pm}+ \xi_{14}^{x^\pm}  \xi_{23}^{x^\pm}
\ea
Using the transformation rule for the   4-point function 
\be\label{ddd}
\braket{1_\pm2_\pm3_\pm4_\pm}^x_\psi=\frac{\partial y^{\pm}_1}{\partial x^\pm_1}\frac{\partial y^\pm_2}{\partial x^\pm_2}\frac{\partial y^\pm_3}{\partial x^\pm_3}\frac{\partial y^\pm_4}{\partial x^\pm_4}\braket{1_\pm2_\pm 3_\pm4_\pm}^y
\ee
Following the same geometric idea used to define the energy momentum tensor, we define the renormalized  4-point correlations function, finite in the coincidence limit, by relating it to the value of the normal ordered   4-point function calculated in the (geometrically defined) local freely falling frame $X^\pm$, namely 
\be\label{covariance}
\braket{1_\pm2_\pm3_\pm4_\pm}^x_{\psi {\rm ren}}=\frac{\partial X^{\pm}_1}{\partial x^\pm_1}\frac{\partial X^\pm_2}{\partial x^\pm_2}\frac{\partial X^\pm_3}{\partial x^\pm_3}\frac{\partial X^\pm_4}{\partial x^\pm_4}\braket{:1_\pm2_\pm3_\pm4_\pm\!:}^X_\psi\equiv J_1J_2J_3J_4 \braket{:1_\pm2_\pm3_\pm4_\pm\!:}^X_\psi, 
\ee
where $J\equiv \frac{\partial X^{\pm}}{\partial x^\pm}$ denotes (from now) the Jacobian of the coordinate transformation linking the coordinates $x^\pm$ and $X^\pm$. It is clear from the previous definition that $\braket{1_\pm2_\pm3_\pm4_\pm}^x_{\psi {\rm ren}}$ is a tensor transforming covariantly under changes of $x^\pm$.

Our task is to express $\braket{1_\pm2_\pm3_\pm4_\pm}^x_{\psi{\rm ren}}$ in terms of normal ordered quantities defined in an arbitrary representation so that we get an expression that is easy to compute with in some situations where a certain `vacuum state' might capture a particular idealized physical situation. 
We then use  \eqref{fff} ---written for the representation constructed in terms of the Minkowskian coordindates $X^\pm$---to write $\braket{:1_\pm2_\pm3_\pm4_\pm\!:}^X_\psi$ in terms of $\braket{1_\pm2_\pm3_\pm4_\pm}^X_\psi$, namely 
\ba
&&\braket{1_\pm2_\pm3_\pm4_\pm}^x_{\psi {\rm ren}} \equiv J_1J_2J_3J_4 \braket{:1_\pm2_\pm3_\pm4_\pm\!:}^X_\psi =J_1J_2J_3J_4 \braket{1_\pm2_\pm3_\pm4_\pm}^X_\psi \n \\
&& - J_1J_2J_3J_4\left(\braket{:1_\pm 2_\pm\!:}^X_\psi \xi_{34}^{X^\pm}+\braket{:1_\pm 3_\pm\!:}^X_\psi \xi_{24}^{X^\pm}+\braket{:1_\pm 4_\pm\!:}^X_\psi \xi_{23}^{X^\pm}+ \right. \n \\ 
&& \left. + \braket{:2_\pm 3_\pm\!:}^X_\psi \xi_{14}^{X^\pm}+\braket{:2_\pm 4_\pm\!:}^X_\psi \xi_{13}^{X^\pm}+\braket{:3_\pm 4_\pm\!:}^X_\psi \xi_{12}^{X^\pm}\right)\n \\
&& - J_1J_2J_3J_4\left(\xi_{12}^{X^\pm}  \xi_{34}^{X^\pm}+ \xi_{13}^{X^\pm}  \xi_{24}^{X^\pm}+ \xi_{14}^{X^\pm}  \xi_{23}^{X^\pm}\right),
\ea
Now equation \eqref{ddd} implies \ba
&&\braket{1_\pm2_\pm3_\pm4_\pm}^x_{\psi {\rm ren}} \equiv J_1J_2J_3J_4 \braket{:1_\pm2_\pm3_\pm4_\pm\!:}^X_\psi =\braket{1_\pm2_\pm3_\pm4_\pm}^x_\psi\n \\
&& - J_1J_2J_3J_4\left(\braket{:1_\pm 2_\pm\!:}^X_\psi \xi_{34}^{X^\pm}+\braket{:1_\pm 3_\pm\!:}^X_\psi \xi_{24}^{X^\pm}+\braket{:1_\pm 4_\pm\!:}^X_\psi\xi_{23}^{X^\pm}
+ \right. \n \\ 
&& \left. + \braket{:2_\pm 3_\pm\!:}^X_\psi \xi_{14}^{X^\pm}+\braket{:2_\pm 4_\pm\!:}^X_\psi \xi_{13}^{X^\pm}+\braket{:3_\pm 4_\pm\!:}^X_\psi \xi_{12}^{X^\pm}\right)\n \\
&& - J_1J_2J_3J_4\left(\xi_{12}^{X^\pm}  \xi_{34}^{X^\pm}+ \xi_{13}^{X^\pm}  \xi_{24}^{X^\pm}+ \xi_{14}^{X^\pm}  \xi_{23}^{X^\pm}\right),
\ea
where we used \eqref{fff}---written for the representation constructed in terms of the Minkowskian coordindates $X^\pm$---to write $\braket{:1_\pm2_\pm3_\pm4_\pm\!:}^X_\psi$ in terms of $\braket{1_\pm2_\pm3_\pm4_\pm}^X_\psi$ followed by equation \eqref{ddd}. We can write the previous equation  as follows
\ba
&&\braket{1_\pm2_\pm3_\pm4_\pm}^x_{\psi {\rm ren}}=\braket{1_\pm2_\pm3_\pm4_\pm}^x_\psi\n \\
&& - J_1J_2J_3J_4\left[(\braket{1_\pm2_\pm}^X_\psi-\xi_{12}^{X^\pm}) \xi_{34}^{X^\pm}+(\braket{1_\pm3_\pm}^X_\psi-\xi_{13}^{X^\pm}) \xi_{24}^{X^\pm}+(\braket{1_\pm4_\pm}^X_\psi-\xi_{14}^{X^\pm}) \xi_{23}^{X^\pm} \right. \n \\ && 
\left. +(\braket{2_\pm3_\pm}^X_\psi -\xi_{23}^{X^\pm}) \xi_{14}^{X^\pm}+(\braket{2_\pm4_\pm}^X_\psi-\xi_{24}^{X^\pm})  \xi_{13}^{X^\pm}+(\braket{3_\pm4_\pm}^X_\psi -\xi_{34}^{X^\pm}) \xi_{12}^{X^\pm}\right]\n \\
&& - J_1J_2J_3J_4\left(\xi_{12}^{X^\pm}  \xi_{34}^{X^\pm}+ \xi_{13}^{X^\pm}  \xi_{34}^{X^\pm}+ \xi_{14}^{X^\pm}  \xi_{23}^{X^\pm}\right),
\ea
using that for instance $J_1J_2\braket{1_\pm2_\pm}^X_\psi=\braket{1_\pm2_\pm}^x_\psi$ and denoting terms like $J_1J_2 \xi_{12}^{X^\pm}=J(\xi_{12}^{X^\pm})$ we obtain
\ba
&&\braket{1_\pm2_\pm3_\pm4_\pm}^x_{\psi {\rm ren}} =\braket{1_\pm2_\pm3_\pm4_\pm}^x_\psi\n \\
&& - \left[(\braket{1_\pm2_\pm}^x_\psi-J(\xi_{12}^{X^\pm})) J(\xi_{34}^{X^\pm})+(\braket{1_\pm3_\pm}^x_\psi-J(\xi_{13}^{X^\pm})) J(\xi_{24}^{X^\pm})+(\braket{1_\pm4_\pm}^x_\psi-J(\xi_{14}^{X^\pm})) J(\xi_{23}^{X^\pm})\right. \n \\ && 
\left. +(\braket{2_\pm3_\pm}^x_\psi -J(\xi_{23}^{X^\pm})) J(\xi_{14}^{X^\pm})+(\braket{2_\pm4_\pm}^x_\psi-J(\xi_{24}^{X^\pm}))  J(\xi_{13}^{X^\pm})+(\braket{3_\pm4_\pm}^x_\psi -J(\xi_{34}^{X^\pm})) J(\xi_{12}^{X^\pm})\right]\n \\
&& - J(\xi_{12}^{X^\pm})  J(\xi_{34}^{X^\pm})- J(\xi_{13}^{X^\pm})  J(\xi_{24}^{X^\pm})- J(\xi_{14}^{X^\pm})  J(\xi_{23}^{X^\pm})\n \\
&&=\braket{1_\pm2_\pm3_\pm4_\pm}^x_\psi\n \\
&& -\braket{1_\pm2_\pm}^x_\psi J(\xi_{34}^{X^\pm})-\braket{1_\pm3_\pm}^x_\psi J(\xi_{24}^{X^\pm})-\braket{1_\pm4_\pm}^x_\psi J(\xi_{23}^{X^\pm})
\n \\ && -\braket{2_\pm3_\pm}^x_\psi J(\xi_{14}^{X^\pm})-\braket{2_\pm4_\pm}^x_\psi J(\xi_{13}^{X^\pm})-\braket{3_\pm4_\pm}^x_\psi J(\xi_{12}^{X^\pm})\n \\
&& + J(\xi_{12}^{X^\pm}) J(\xi_{34}^{X^\pm})+J(\xi_{13}^{X^\pm}) J(\xi_{24}^{X^\pm})+J(\xi_{14}^{X^\pm}) J(\xi_{23}^{X^\pm}) +J(\xi_{23}^{X^\pm}) J(\xi_{14}^{X^\pm})+J(\xi_{24}^{X^\pm})  J(\xi_{13}^{X^\pm})+J(\xi_{34}^{X^\pm}) J(\xi_{12}^{X^\pm})\n \\
&& - J(\xi_{12}^{X^\pm})  J(\xi_{34}^{X^\pm})- J(\xi_{13}^{X^\pm})  J(\xi_{24}^{X^\pm})- J(\xi_{14}^{X^\pm})  J(\xi_{23}^{X^\pm})\n \\
&&=\braket{1_\pm2_\pm3_\pm4_\pm}^x_\psi\n \\
&& -\braket{1_\pm2_\pm}^x_\psi J(\xi_{34}^{X^\pm})-\braket{1_\pm3_\pm}^x_\psi J(\xi_{24}^{X^\pm})-\braket{1_\pm4_\pm}^x_\psi J(\xi_{23}^{X^\pm})
\n \\
&& -\braket{2_\pm3_\pm}^x_\psi J(\xi_{14}^{X^\pm})-\braket{2_\pm4_\pm}^x_\psi J(\xi_{13}^{X^\pm})-\braket{3_\pm4_\pm}^x_\psi J(\xi_{12}^{X^\pm})\n \\
&& + J(\xi_{12}^{X^\pm}) J(\xi_{34}^{X^\pm})+J(\xi_{13}^{X^\pm}) J(\xi_{24}^{X^\pm})+J(\xi_{14}^{X^\pm}) J(\xi_{23}^{X^\pm})
\ea
Using again \eqref{fff} to express $\braket{1_\pm2_\pm3_\pm4_\pm}^x_\psi$ in terms of normal ordered quantities with respect to $\ket{0_x}$ and (for example) $\braket{1_\pm2_\pm}^x_\psi=\braket{:1_\pm 2_\pm\!:}^x_\psi +\xi_{12}^{x^\pm}$ we get
\ba
&& \braket{1_\pm2_\pm3_\pm4_\pm}^x_{\psi {\rm ren}} =\braket{:1_\pm2_\pm3_\pm4_\pm\!:}^x_\psi\\ && +\braket{:1_\pm 2_\pm\!:}^x_\psi \xi_{34}^{x^\pm}+\braket{:1_\pm 3_\pm\!:}^x_\psi \xi_{24}^{x^\pm}+\braket{:1_\pm 4_\pm\!:}^x_\psi \xi_{23}^{x^\pm}+\braket{:2_\pm 3_\pm\!:}^x_\psi \xi_{14}^{x^\pm}+\braket{:2_\pm 4_\pm\!:}^x_\psi \xi_{13}^x +\braket{:3_\pm 4_\pm\!:}^x_\psi \xi_{12}^{x^\pm}+\n \\
&&  +\xi_{12}^{x^\pm}  \xi_{34}^{x^\pm}+ \xi_{13}^x  \xi_{24}^{x^\pm}+ \xi_{14}^{x^\pm}  \xi_{23}^{x^\pm}\n\\
&& -(\braket{:1_\pm 2_\pm\!:}^x_\psi+\xi_{12}^{x^\pm})J(\xi_{34}^{X^\pm})-(\braket{:1_\pm 3_\pm\!:}^x_\psi+\xi_{13}^x) J(\xi_{24}^{X^\pm})-(\braket{:1_\pm 4_\pm\!:}^x_\psi+\xi_{14}^{x^\pm}) J(\xi_{23}^{X^\pm})\n \\
&&-(\braket{:2_\pm3_\pm\!:}^x_\psi+\xi_{23}^{x^\pm})J(\xi_{14}^{X^\pm})-(\braket{:2_\pm4_\pm\!:}^x_\psi+\xi_{24}^{x^\pm}) J(\xi_{13}^{X^\pm})-(\braket{:3_\pm4_\pm\!:}^x_\psi+\xi_{34}^{x^\pm}) J(\xi_{12}^{X^\pm})\n \\
&& + J(\xi_{12}^{X^\pm}) J(\xi_{34}^{X^\pm})+J(\xi_{13}^{X^\pm}) J(\xi_{24}^{X^\pm})+J(\xi_{14}^{X^\pm}) J(\xi_{23}^{X^\pm})\n \\ 
&& =\braket{:1_\pm2_\pm3_\pm4_\pm\!:}^x_\psi \n\\ 
&&+ \braket{:1_\pm 2_\pm\!:}^x_\psi \left[\xi_{34}^{x^\pm}-J(\xi_{34}^{X^\pm})\right]+\braket{:1_\pm 3_\pm\!:}^x_\psi [\xi_{24}^{x^\pm}-J(\xi_{24}^{X^\pm})]+\braket{:1_\pm 4_\pm\!:}^x_\psi[\xi_{23}^{x^\pm}-J(\xi_{23}^{X^\pm})] \n \\ 
&& +\braket{:2_\pm 3_\pm\!:}^x_\psi [\xi_{14}^{x^\pm}-J(\xi_{14}^{X^\pm})]+\braket{:2_\pm 4_\pm\!:}^x_\psi [\xi_{13}^{x^\pm}-J(\xi_{13}^{X^\pm})]+\braket{:3_\pm 4_\pm\!:}^x_\psi [\xi_{12}^{x^\pm}-J(\xi_{12}^{X^\pm})]+\n \\
&&  +\xi_{12}^{x^\pm}  \xi_{34}^{x^\pm}+ \xi_{13}^x  \xi_{24}^{x^\pm}+ \xi_{14}^{x^\pm}  \xi_{23}^{x^\pm}\n\\
&& -\xi_{12}^{x^\pm} J(\xi_{34}^{X^\pm})-\xi_{13}^x J(\xi_{24}^{X^\pm})-\xi_{14}^{x^\pm} J(\xi_{23}^{X^\pm})-\xi_{23}^{x^\pm}J(\xi_{14}^{X^\pm})-\xi_{24}^{x^\pm} J(\xi_{13}^{X^\pm})-\xi_{34}^{x^\pm} J(\xi_{12}^{X^\pm})\n \\
&& + J(\xi_{12}^{X^\pm}) J(\xi_{34}^{X^\pm})+J(\xi_{13}^{X^\pm}) J(\xi_{24}^{X^\pm})+J(\xi_{14}^{X^\pm}) J(\xi_{23}^{X^\pm})\n \\
&& =\braket{:1_\pm2_\pm3_\pm4_\pm\!:}^x_\psi +\braket{:1_\pm 2_\pm\!:}^x_\psi  \left[\xi_{34}^{x^\pm}-J(\xi_{34}^{X^\pm})\right]+\braket{:1_\pm 3_\pm\!:}^x_\psi  [\xi_{24}^{x^\pm}-J(\xi_{24}^{X^\pm})]+\braket{:1_\pm 4_\pm\!:}^x_\psi [\xi_{23}^{x^\pm}-J(\xi_{23}^{X^\pm})] +
\n \\ 
&&
 +\braket{:2_\pm 3_\pm\!:}^x_\psi  [\xi_{14}^{x^\pm}-J(\xi_{14}^{X^\pm})]+\braket{:2_\pm 4_\pm\!:}^x_\psi  [\xi_{13}^x-J(\xi_{13}^{X^\pm})]+\braket{:3_\pm 4_\pm\!:}^x_\psi  [\xi_{12}^{x^\pm}-J(\xi_{12}^{X^\pm})]\n \\ 
 &&+[\xi_{12}^{x^\pm}-J(\xi_{12}^{X^\pm})] [\xi_{34}^{x^\pm}-J(\xi_{34}^{X^\pm})]+ [\xi_{13}^x-J(\xi_{13}^{X^\pm})] [\xi_{24}^{x^\pm}-J(\xi_{24}^{X^\pm})]+ [\xi_{14}^{x^\pm}-J(\xi_{14}^{X^\pm})] [\xi_{23}^{x^\pm}-J(\xi_{23}^{X^\pm})]\n
\ea
Writing the last equality  separately
\ba\label{lalila}
&& \braket{1_\pm2_\pm3_\pm4_\pm}^x_{\psi{\rm ren}} = \braket{:1_\pm2_\pm3_\pm4_\pm\!:}^x_\psi +\braket{:1_\pm 2_\pm\!:}^x_\psi  [\xi_{34}^{x^\pm}-J(\xi_{34}^{X^\pm})]+\braket{:1_\pm 3_\pm\!:}^x_\psi  [\xi_{24}^{x^\pm}-J(\xi_{24}^{X^\pm})] \\ &&  \n +\braket{:1_\pm 4_\pm\!:}^x_\psi [\xi_{23}^{x^\pm}-J(\xi_{23}^{X^\pm})] +\braket{:2_\pm 3_\pm\!:}^x_\psi  [\xi_{14}^{x^\pm}-J(\xi_{14}^{X^\pm})] \n \\ && +\braket{:2_\pm 4_\pm\!:}^x_\psi  [\xi_{13}^x-J(\xi_{13}^{X^\pm})]+\braket{:3_\pm 4_\pm\!:}^x_\psi  [\xi_{12}^{x^\pm}-J(\xi_{12}^{X^\pm})]+ \n \\
&& +[\xi_{12}^{x^\pm}-J(\xi_{12}^{X^\pm})] [\xi_{34}^{x^\pm}-J(\xi_{34}^{X^\pm})]+ [\xi_{13}^x-J(\xi_{13}^{X^\pm})] [\xi_{24}^{x^\pm}-J(\xi_{24}^{X^\pm})]+ [\xi_{14}^{x^\pm}-J(\xi_{14}^{X^\pm})] [\xi_{23}^{x^\pm}-J(\xi_{23}^{X^\pm})]\n
\ea
Using that the quantities in the square brakets all have a well defined coincidence limit given by 
\be
\lim_{1\to 2}\, [\xi_{12}^{x^\pm}-J(\xi_{12}^{X^\pm})]=-\frac{\hbar}{24\pi}\{X^{\pm},x^{\pm}\}.
\ee
Note that all quanties are well defined in the coincidence limit so that by taking it we get the desired result:
\ba
&& \braket{\psi|\partial_\pm\phi(x)\partial_\pm\phi(x)\partial_\pm\phi(x)\partial_\pm\phi(x)|\psi}_{\rm ren} =\n \\ 
&&=\braket{\psi|:\partial_\pm\phi(x)\partial_\pm\phi(x)\partial_\pm\phi(x)\partial_\pm\phi(x):|\psi}-\frac{\hbar}{4\pi}\braket{\psi|:\partial_\pm\phi(x)\partial_\pm\phi(x):\psi}\{X,x\}+ \frac{\hbar^2}{192\pi^2} \{X,x\}^2
\ea
Finally, using \eqref{swm} we get the following expression of the renormalized expectation value of the square of the energy momentum tensor:
\ba\label{ckey}
\braket{\psi|T_{\pm \pm}T_{\pm \pm}|\psi}_{\rm ren}&\equiv& \braket{\psi|:T_{\pm \pm}T_{\pm \pm}:|\psi} \\ &+&\frac{\hbar}{2\pi}\braket{\psi|:T_{\pm \pm}:|\psi}\left(\partial_{\pm}\log(\Omega)\partial_{\pm}\log(\Omega)-\partial^2_{\pm}\log(\Omega)\right) + \frac{\hbar^2}{48 \pi} \left(\partial_{\pm}\log(\Omega)\partial_{\pm}\log(\Omega)-\partial^2_{\pm}\log(\Omega)\right)^2,\n 
\ea
whose covariance is granted by construction from the definition \eqref{covariance}. The same method allows to write the $n$-point correlation functions for arbitrary $n$ in a closed manner. 

\subsection{Calculation of $\braket{\psi|T_{++}T_{--}|\psi}_{\rm ren}$}

In this case there are many terms that vanish in the Wick expansion of the relevant 4-point function due to the fact that $\braket{0_x|\partial_+\phi(x_1)\partial_-\phi(x_2)|0_x}=0$. We obtain
\ba
\braket{1_+2_+3_-4_-}^x_\psi  &=& \braket{:1_+2_+3_-4_-\!:}^x_\psi  +\braket{:1_+ 2_+\!:}^x_\psi  \xi_{34}^{x^-}+\braket{:3_- 4_-\!:}^x_\psi  \xi_{12}^{x^+}
  +\xi_{12}^{x^+}  \xi_{34}^{x^-}.\ea
The renormalization prescription leads to
\ba
&&\braket{1_+2_+3_-4_-}^x_{\psi {\rm ren}} \equiv J_1J_2J_3J_4 \braket{:1_+2_+3_-4_-\!:}^X_\psi = \\ && \n =\braket{1_+2_+3_-4_-}^x_\psi 
 - J_1J_2J_3J_4\left(\braket{:1_+ 2_+\!:}^X_\psi  \xi_{34}^{X^-}+\braket{:3_- 4_-\!:}^X_\psi  \xi_{12}^{X^+}+\xi_{12}^{X^+}  \xi_{34}^{X^-}\right)
=\\ && =\braket{1_+2_+3_-4_-}^x_\psi 
 - J_1J_2J_3J_4\left(\braket{1_+ 2_+}^X_\psi  \xi_{34}^{X^-}+\braket{3_- 4_-}^X_\psi  \xi_{12}^{X^+}+\xi_{12}^{X^+}  \xi_{34}^{X^-}\right)+J_1J_2J_3J_4\left(\xi_{12}^{X^+} \xi_{34}^{X^-}+\xi_{34}^{X^-} \xi_{12}^{X^+}\right)\n=\\ &&\n  =\braket{1_+2_+3_-4_-}^x_\psi 
 -\braket{1_+ 2_+}^x_\psi  J(\xi_{34}^{X^-})-\braket{3_- 4_-}^x_\psi  J(\xi_{12}^{X^+})+J(\xi_{12}^{X^+}) J(\xi_{34}^{X^-})
=\\ && =\braket{:1_+2_+3_-4_-\!:}^x_\psi  +\braket{:1_+ 2_+\!:}^x_\psi  \xi_{34}^{x^-}+\braket{:3_- 4_-\!:}^x_\psi  \xi_{12}^{x^+}
  +\xi_{12}^{x^+}  \xi_{34}^{x^-} \n \\ && 
 -\braket{:1_+ 2_+:}^x_\psi  J(\xi_{34}^{X^-})-\braket{:3_- 4_-:}^x_\psi  J(\xi_{12}^{X^+}) -\xi_{12}^{x^+} J(\xi_{34}^{X^-})- \xi_{34}^{x^-}J(\xi_{12}^{X^+})+J(\xi_{12}^{X^+}) J(\xi_{34}^{X^-})\n 
=\\ && =\braket{:1_+2_+3_-4_-\!:}^x_\psi  +\braket{:1_+ 2_+\!:}^x_\psi  (\xi_{34}^{x^-}-J(\xi_{34}^{X^-})+\braket{:3_- 4_-\!:}^x_\psi  (\xi_{12}^{x^+}-J(\xi_{12}^{X^+})
  +  (\xi_{12}^{x^+}-J(\xi_{12}^{X^+})(\xi_{34}^{x^-}-J(\xi_{34}^{X^-}),\n 
\ea
where we have followed step by step what we did previously. Thus,
\ba
&& \braket{\psi|\partial_+\phi(x)\partial_+\phi(x)\partial_-\phi(x)\partial_-\phi(x)|\psi}_{\rm ren} =\n \\ 
&&=\braket{\psi|:\partial_+\phi(x)\partial_+\phi(x)\partial_-\phi(x)\partial_-\phi(x):|\psi}-\frac{\hbar}{24\pi}\braket{\psi|:\partial_+\phi(x)\partial_+\phi(x):\psi}\{X^-,x^-\}\n \\ && -\frac{\hbar}{24\pi}\braket{\psi|:\partial_-\phi(x)\partial_-\phi(x):\psi}\{X^+,x^+\}+ \frac{\hbar^2}{576\pi^2} \{X^+,x^+\}\{X^-,x^-\}.
\ea
And finally, using \eqref{swm}, we get
\ba
&& \braket{\psi|T_{++}T_{--}|\psi}_{\rm ren} =\braket{\psi|:T_{++}T_{--}:|\psi}\n\\ 
&& +\frac{\hbar}{12\pi}\braket{\psi|:T_{++}:\psi}\left(\partial_{-}\log(\Omega)\partial_{-}\log(\Omega)-\partial^2_{-}\log(\Omega)\right)
 +\frac{\hbar}{12\pi}\braket{\psi|:T_{--}:\psi}\left(\partial_{+}\log(\Omega)\partial_{+}\log(\Omega)-\partial^2_{+}\log(\Omega)\right)\n \\ && + \frac{\hbar^2}{144\pi^2}\left(\partial_{-}\log(\Omega)\partial_{-}\log(\Omega)-\partial^2_{-}\log(\Omega)\right)\left(\partial_{+}\log(\Omega)\partial_{+}\log(\Omega)-\partial^2_{+}\log(\Omega)\right).
\ea
Applying the previous equation to the vacuum states associated with the notion of normal ordering involved, 
\ba
&& \braket{0|T_{++}T_{--}|0}_{\rm ren}= \frac{\hbar^2}{144\pi^2}\left(\partial_{-}\log(\Omega)\partial_{-}\log(\Omega)-\partial^2_{-}\log(\Omega)\right)\left(\partial_{+}\log(\Omega)\partial_{+}\log(\Omega)-\partial^2_{+}\log(\Omega)\right)\n \\
&&=\braket{0|T_{++}|0}_{\rm ren}\braket{0|T_{--}|0}_{\rm ren}
\ea
We see that the large `vacuum fluctuations' are not going to manifest in such correlation functions. The reason is that $[T_{++},T_{--}]=0$ and the conformal vacuum states are product states.  

\subsection{Operator product expansion}\label{OPE}

One can express the previous results in the language of the OPE. For the energy momentum tensor, the relevant expansion is
\ba
\braket{\psi|\partial_{x^\pm}\phi_1 \partial_{x^\pm} \phi_2 |\psi}&=&J_1J_2 \braket{\psi|\partial_{X^\pm}\phi_1 \partial_{X^\pm} \phi_2 |\psi}\n \\ &\approx& J_1J_2 \left(\braket{\psi|:\partial_{X^\pm}\phi_1\partial_{X^\pm}\phi_2\!:_X|\psi}
\underbrace{-\frac{\hbar}{12\pi}\partial_{X^\pm_1}\partial_{X^\pm_2} \log\left({|(X^+_1-X^+_2)(X^-_1-X^-_2)|}\right)}_{\braket{0_X|\partial_\pm\phi(X_1)\partial_\pm\phi(X_2)|0_X}=-\frac{\hbar}{12\pi} \frac{1}{|X^\pm_1-X^\pm_2|^2}}\right)
\n \\
&\approx& \braket{\psi|(\partial_{x^\pm}\phi)^2|\psi}_{\rm ren}-\frac{\hbar}{12\pi}\partial_{x_1^\pm}\partial_{x_2^\pm} \log\left({\sigma(x_1,x_2)},\right)
\ea
where $\sigma(x_1,x_2)$ is the geodesic distance squared between $x_1$ and $x_2$, and $\approx$ denotes the equality when $x_1\to x_2$.
Similarly
\ba
\braket{\psi|\partial_{x^\pm}\phi_1 \partial_{x^\pm} \phi_2 \partial_{x^\pm} \phi_3 \partial_{x^\pm} \phi_4 |\psi}&\approx& \braket{\psi|(\partial_{x^\pm}\phi)^4|\psi}_{\rm ren}  \\ &-& \n \frac{\hbar}{2\pi} \braket{\psi|(\partial_{x^\pm}\phi)^2|\psi}_{\rm ren}\partial_{x_1^\pm}\partial_{x_2^\pm} \log\left({\sigma(x_1,x_2)}\right)\\ &+& \n \frac{\hbar^2}{48\pi} \partial_{x_1^\pm}\partial_{x_2^\pm} \partial_{x_3^\pm}\partial_{x_4^\pm} \log\left({\sigma(x_1,x_2)}\right)\log\left({\sigma(x_3,x_4)}\right).
\ea
One can express the OPE for any correlation function of primary fields along the same lines.

\section{Relation to 4d}\label{apalache}

The Klein-Gordon equation $\square\Phi-m^2\Phi=0$ in the Schwarzschild background  in spherical coordinates can be written 
as
\be
4\partial_u\partial_v \phi_\ell+\underbrace{\left(1-\frac{2M}{r}\right)\left(\frac{\ell(\ell+1)}{r^2}+\frac{2M}{r^3}+m^2\right)}_{V_\ell(r)}\phi_\ell=0,
\ee
where we used the ansatz $\Phi(u,v,\theta, \varphi)=\phi_\ell(u,v) Y_{\ell m}(\theta, \varphi)/r$ with $u=t-r_*$, $v=t+r_*$, 
$r_*=r+2M\log(r/(2M) -1)$ and the radial coordinate should be taken as an implicit function of the null coordinates, $r=r(u,v)$.
At the horizon $r=2M$ the previous equation becomes the 2d conformally invariant field equation of a massless scalar field
for all the modes, namely 
\be
\partial_u\partial_v \phi_\ell=0.
\ee 
The non locality of quantum field theory implies that the effects of the potential $V_\ell(r)$
cannot be neglected in discussion involving the Hilbert space construction and the definition of some particular states of interest.
However, qualitative information about the quantum theory can be obtained especially if we simplify the discussion further by focusing on the spherically symmetric modes $\ell=0$ (which is the main channel for scalar Hawking radiation \cite{Page:1976df}). Identifying $\phi_0$ with the 2d scalar field $\phi$ of Section \ref{2d} we have that the contribution of these modes to the relevant components of the energy momentum tensor of the Klein-Gordon field is
\be
\left.T^{\rm sph}_{vv}\right|_{\rm H}=\left.\partial_v\Phi\partial_v\Phi\right|_{\rm H}=\left.\left(\frac{\partial_v\phi}{r}-\frac{\phi}{r^2} \frac{\partial r}{\partial v}\right)\left(\frac{\partial_v\phi}{r}-\frac{\phi}{r^2} \frac{\partial r}{\partial v}\right)\right|_{\rm H}=\left.\frac{\partial_v\phi\partial_v\phi}{4M^2}\right|_{\rm H}=\frac{T^{\rm 2d}_{vv}}{4M^2}.
\ee

   \section{Squeezing is a relative notion}\label{Squezzed}
   
 It is  well known that in quantum filed theory  and for generic  space-times there is no canonical  notion of a vacuum  state.   In certain situations (for instance when the space-time  possesses a time translation symmetry  reflected  by  the existence of a time-like  killing field) one can use the  associated  notion of positive frequency to characterize  the  modes  of the field,    and  use that notion to define a preferred vacuum state. Concretely, in the case of a scalar field (used here for illustration) one defines the field operator acting in the Fock representation via the sum
 \be\phi(x) =\sum_i a_{i} f_i(x)+a^{\dagger}_i \bar f_i.\ee
where $\{f_i\}$ is a basis of positive frequency solutions of the field equations and the coefficients $a_{i}$ and $a_i^\dagger$ are annihilation and creation operators operators. The associated vacuum state $ |0 \rangle$  is defined by the  condition $a_{i} |0 \rangle =0$.  One can define a  related  state $  | {\tilde 0 }  \rangle $ that is  considered as ``squezzed"  with respect  to the former  by considering  a Bogoliubov  transformation $ b_{j} = \alpha_{j}^i  a_{i} + \beta^{i}_j a^{\dagger}_{i} $ with  $\boldsymbol{\alpha} ^\dagger \boldsymbol{\alpha} -\boldsymbol{\beta} ^\dagger \boldsymbol{\beta}= {\bf 1}$ and the demand  that 
     $ b_i | {\tilde 0}  \rangle =0$.  It  is clear that squezzing is  then a relative notion, with  the  state   
     $  | {\tilde 0} \rangle =0$ being  squezzed  w. r. t.   $ |0 \rangle$ and the latter  being  squezzed  w. r. t. the former. Both,  the vaccum states  and  the  related squezzed states have the property that the expectation value of the  field  vanishes.   The renormalized  energy momentum tensor in such states  needs not  vanish (as  is  well known  say for the  type  of states that occur in  the  study of black hole  evaporation) illustrating the fact that,  generically,  such states  differ locally from the  so called ``freely falling vacuum" ( that is the state constructed using local Riemann  coordinates  as we have explicitly done here 2d \cite{Fabbri:2005mw}).  It is for   those cases that  according to  our  explorations the quantum uncertainties of the  energy momentum tensor are as  large as the energy momentum tensor  expectation value itself, and thus  would  fail, according to our proposed  criterion,  to  provide  suitable states for  semi-classical gravity.

\end{appendix}

\end{document}